\documentclass[iop,apj]{emulateapj}
\usepackage{apjfonts}
\usepackage{graphicx}
\usepackage{color}
\usepackage{amsmath}
\newcommand{\dlmdla}{\ensuremath{d\ln{M}/d\ln{A}}}
\newcommand{\phpm}{\mbox{\ensuremath{\phantom{-}}}}
\newcommand{\hefour}{HE\,0435$-$1223}
\newcommand{\err}[2]{\ensuremath{^{_{+#1}}_{^{-#2}}}}
\begin{document}
\bibliographystyle{apj}
\shorttitle{EFFECT OF A TIME-VARYING ACCRETION DISK SIZE}
\shortauthors{BLACKBURNE \& KOCHANEK}
\slugcomment{}
\title{The Effect of a Time-Varying Accretion Disk Size on Quasar Microlensing\\
  Light Curves}

\author{Jeffrey~A.~Blackburne \& Christopher~S.~Kochanek}
\affil{\scriptsize{Department of Astronomy and Center for Cosmology and AstroParticle Physics,
  The Ohio State University, 140 West 18th Avenue, Columbus, OH 43210, USA;\\
  blackburne@astronomy.ohio-state.edu, ckochanek@astronomy.ohio-state.edu}}
\begin{abstract}

Microlensing perturbations to the magnification of gravitationally
lensed quasar images are dependent on the angular size of the
quasar. If quasar variability at visible wavelengths is caused by a
change in the area of the accretion disk, it will affect the
microlensing magnification. We derive the expected signal, assuming
that the luminosity scales with some power of the disk area, and
estimate its amplitude using simulations. We discuss the prospects for
detecting the effect in real-world data and for using it to estimate
the logarithmic slope of the luminosity's dependence on disk
area. Such an estimate would provide a direct test of the standard
thin accretion disk model. We tried fitting six seasons of the light
curves of the lensed quasar \hefour\ including this effect as a
modification to the Kochanek et al. (2006) approach to estimating time
delays.  We find a dramatic improvement in the goodness of fit and
relatively plausible parameters, but a robust estimate will require a
full numerical calculation in order to correctly model the strong
correlations between the structure of the microlensing magnification
patterns and the magnitude of the effect. We also comment briefly on
the effect of this phenomenon for the stability of time delay
estimates.

\end{abstract}


\section{Introduction}
\label{sec:intro}

Recent observational studies of the microlensing of gravitationally
lensed quasars \citep[see review by][]{Wambsganss:2006p453} have led
to new constraints on the structure of the innermost regions of
quasars and on the properties of foreground galaxies. Because quasar
accretion disks subtend angles comparable to the Einstein radii of
stars in foreground galaxies, studies taking advantage of the
dependence of microlensing magnifications on the source size have been
able to estimate the sizes of accretion disks \citep{Pooley:2007p19,
Anguita:2008p615, Dai:2010p278, Morgan:2010p0}, and multi-wavelength
observations have enabled estimates of the wavelength dependence of
the disk size \citep{Poindexter:2008p34, Eigenbrod:2008p933,
Bate:2008p1955, Floyd:2009p233, Mosquera:2009p1292}. Other studies
have estimated the ratio of clumpy stellar mass to more smoothly
distributed dark matter in lensing galaxies, providing the only
constraints on stellar mass fractions not dependent on the initial
mass function \citep[e.g.][]{Pooley:2009p1892}. Time domain
observations of microlensing variability are a particularly powerful
tool for this type of investigation. For example, they have been used
recently to estimate the inclination of a quasar accretion disk
\citep{Poindexter:2009p3669} as well as its transverse velocity
relative to the lens and the mean mass of the stars in the lensing
galaxy \citep{Poindexter:2009p3213}.

To date, quasar microlensing has nearly always been modeled using
accretion disks with time-independent sizes; this ought to be a good
first-order approximation. But the fact that quasar luminosities are
time-variable indicates that this model cannot hold in
detail. Observationally, short-term variations in individual image
fluxes have been observed for a few lensed quasars
\citep{Schild:1996p125, Kundic:1997p75, Schechter:2003p657}, leading
to hypotheses involving microlensing of disks with fast-moving hot
spots \citep{Gould:1997pL13, Schechter:2003p657} or obscured by
fast-moving optically thick broad-line clouds
\citep{Wyithe:2002p615}.

In the spirit of these studies, we investigate a model of a
time-varying accretion disk with a size proportional to some power of
the quasar luminosity. This model introduces one new parameter, which
is the exponent $\epsilon$ between the disk effective area and
luminosity (i.e. $A \propto L^\epsilon$); for simple blackbody disks
they are directly proportional ($\epsilon = 1$). We show that
observations of the light curves of multiply imaged quasars have the
potential to detect this disk size variability, and discuss how such
detections may be used to constrain the value of $\epsilon$. In
Section~\ref{sec:math} we outline the mathematical model for the
observed light curve of a quasar image in terms of the intrinsic
quasar light curve. In Section~\ref{sec:simulations} we study the
amplitude of the effect for a time-variable disk by simulating
realistic microlensing scenarios. We describe in
Section~\ref{sec:epsilon} an algorithm for measuring the effect in
real-world light curves of lensed quasars, and report in
Section~\ref{sec:hefour} the results of applying it to the light
curves of the quasar \hefour. In Section~\ref{sec:conclusions} we give
a brief overview of our conclusions.

\section{Mathematical model}
\label{sec:math}

The optical and near-infrared continuum luminosity of bright quasars is
thought to arise in a thin accretion disk radiating as a
multi-temperature blackbody \citep[see review by][]{Blaes:2007p75}. If
the blackbody effective temperature varies as a power law with radius,
$T_\mathrm{eff} = T_0 (R/R_0)^{-\beta}$, then by integrating the
Planck function over all radii, the total luminosity of the disk can
be shown to be proportional to $R_0^2 K(\beta)$, where $K(\beta)$ is a
dimensionless function of the power law slope. This calculation has
been used as a means of estimating the disk size from the optical
luminosity \citep[e.g.][]{Morgan:2010p0}, but our main concern is
the scaling. The luminosity of a blackbody disk with a power-law
temperature profile will scale linearly with its effective area
$R_0^2$.

This scaling implies that quasar variability at ultraviolet and optical
rest wavelengths is matched by a proportional variability in the size
of the accretion disk. This is the simplest model for a time-variable
disk size, because it does not require the introduction of
substructure in the disk or any new time scale. In the case of quasars
that are experiencing microlensing, the source size dependence of the
microlensing magnification will affect an image's light curve in phase
with the quasar variability, boosting or suppressing it depending on
the details of the microlens configuration.

To quantify the effect, we start with the observed flux of a lensed
quasar image as a function of time,
\begin{equation}
f(t) = L(t) M  + f_C ~,
\end{equation}
where $L(t)$ is the time-dependent quasar luminosity and $M$ is the
effective magnification --- specifically, the product of the
macro-magnification and the microlensing magnification. The latter can
change with time, but for this first discussion we will assume that it
is constant on the time scales of quasar variability. $M$ does,
however, change with the area of the disk, because the microlensing
magnification pattern is the convolution of the true caustic pattern
with the (changing) source profile. The $f_C$ term denotes
contaminating flux from non-disk sources, such as broad emission
lines, the quasar host galaxy or the lens galaxy. The flux (and
likewise the quasar luminosity $L$ and disk area $A$) may be
decomposed into a DC component and a small time-variable component:
\begin{equation}
f(t) = \langle f \rangle + \delta f(t) ~,
\end{equation}
where
\begin{align}
\langle f \rangle &= \langle ML \rangle (1+C) ~,~ \mathrm{ and} \notag \\
\delta f(t) &= \langle M \rangle \delta L(t) + \langle L \rangle \frac{dM}{dA} \delta
A(t) ~.
\end{align}
Here $C \equiv f_C/\langle ML \rangle$ is the fractional flux
contamination. Thus,
\begin{align}
\label{eqn:fluxperturb}
\frac{\delta f}{\langle f \rangle} =& \notag
\left(\frac{\delta L}{\langle L \rangle} + \frac{d\ln{M}}{d\ln{A}}
\frac{\delta A}{\langle A \rangle}\right)\left(1 + C\right)^{-1} \\
\approx& \frac{\delta L}{\langle L \rangle}
\left(1+\epsilon \frac{d\ln{M}}{d\ln{A}}-C\right) ~,
\end{align}
where $\epsilon \equiv (d\ln{L}/d\ln{A})^{-1}$ is the logarithmic
change in disk luminosity with disk area. If the luminosity of the
disk is simply proportional to its area, then $\epsilon = 1$. We have
assumed in the second line that $C \ll 1$, and that \dlmdla\ is also
small (we verify this assumption in Section~\ref{sec:simulations}). To
convert this expression to magnitudes, we first separate the light
curve into a time-averaged component $\langle m \rangle = -2.5 \log
\langle f \rangle$ and a small variable component $\delta m(t) \propto
\delta f / \langle f \rangle$. Thus
\begin{equation}
\label{eqn:magperturb}
\delta m(t) \approx \delta m_\mathrm{int}(t) 
\left(1+\epsilon \frac{d\ln{M}}{d\ln{A}}-C\right) ~,
\end{equation}
where $\delta m_\mathrm{int}(t) \propto \delta L / \langle L \rangle$
is the intrinsic time-variable light curve of the quasar. This
intrinsic variability is identical for all the images of a lensed
quasar (once lensing time delays have been accounted for). Thus, for a
pair of images with low amounts of contaminating flux, a difference in
\dlmdla\ will cause a difference in the amplitude of quasar
variability seen in the two images. If the former is larger than a few
percent, the latter ought to be detectable when comparing measured
light curves. Observationally, for quasar images labeled A and B, the
intrinsic light curve measured in image B, relative to that measured
in A, would seem to be multiplied by the function
\begin{equation}
\label{eqn:magdiff}
\zeta(t) = 1 + \epsilon\left[\left(\frac{d\ln{M}}{d\ln{A}}\right)_B -
\left(\frac{d\ln{M}}{d\ln{A}}\right)_A\right] - C_B + C_A ~.
\end{equation}

\section{Simulations}
\label{sec:simulations}

We calculated the probability distribution functions of the value of
\dlmdla\ for several realistic microlensing parameter values. We also
created an example of a simulated light curve that includes the effect
of a time-varying disk, in order to test the validity of the
assumptions we made in Section~\ref{sec:math}.

\begin{figure}
  \centering
  \includegraphics[width=0.45\textwidth]{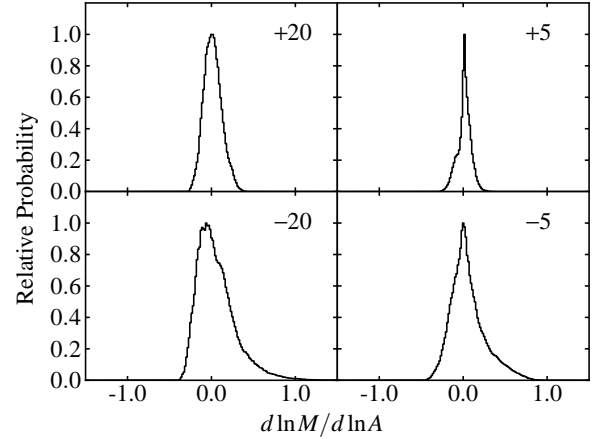}
  \caption{Examples of the distributions of \dlmdla\ for a selection
    of macro-magnifications (shown at the upper right of each
    plot). The source half-light radius is fixed at $10^{-0.5} = 32$\%
    of a solar mass Einstein radius, and the stellar mass fraction is
    fixed at 10\%.}
  \label{fig:histograms}
\end{figure}

For these simulations, we used magnification maps created using the
inverse ray-shooting software of \citet{Wambsganss:1990p407}. The maps
had a total surface mass density (or convergence)
$\kappa_\mathrm{tot}$ roughly equal to their shear $\gamma$,
divided into contributions from smooth matter and point masses. We
specify a broken power-law mass function for the microlens
stars. Between $0.08\,M_\odot$ and $0.5\,M_\odot$, its logarithmic
slope is $-1.8$, and above $0.5\,M_\odot$ it steepens to $-2.7$. This
mass function is very similar to that of \citet{Kroupa:2001p231}. We
cut off the mass function at $1.5\,M_\odot$, because lens galaxies are
typically early-type galaxies with old stellar populations. With this
mass function, the average microlens mass is $\langle m \rangle =
0.247\,M_\odot$, in reasonable agreement with the recent measurement
by \citet{Poindexter:2009p3213}.

Each map encodes the deviation in the magnification of its quasar
image from that produced by a smooth mass distribution as a function
of the position of the source. They are $2000$ pixels on a side; this
is $20$ times the Einstein radius of a solar-mass microlens star. When
projected back to the quasar, the side length is $\sim 5 \times
10^{17}$ cm (the exact number depends on the redshifts of the lens and
the quasar), and the pixel size is thus $\sim 2.5 \times 10^{14}$ cm,
or a few gravitational radii for a $10^9\,M_\odot$ black hole. This is
much smaller than the size of the optical accretion disk.

\subsection{Histograms}
\label{sec:hist}

For any particular quasar image, \dlmdla\ depends on the microlensing
caustic pattern associated with that image and the size of the
accretion disk. One can calculate a probability distribution for its
value by first taking the difference of two copies of an appropriate
pattern which have been convolved with source profiles of slightly
different areas, dividing the result by the difference in the
logarithm of those areas, and finally creating a histogram of the
resulting pattern.

\begin{figure}
  \centering
  \includegraphics[width=0.45\textwidth]{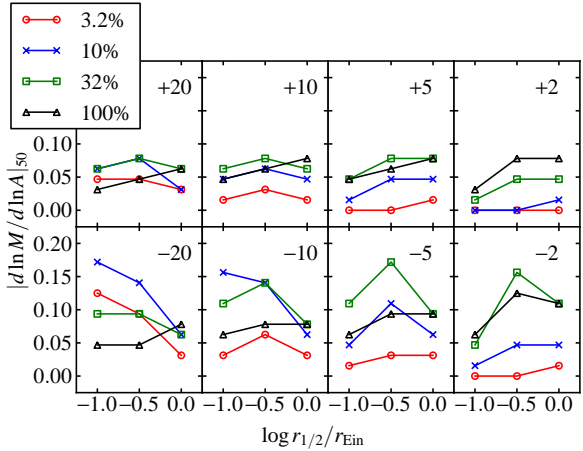}
  \caption{50\% probability bounds on $|\dlmdla|$ as a function of the
  ratio of the source half-light radius to a solar-mass Einstein
  radius. Macro-magnifications are shown in the upper right of each
  plot. The surface mass fraction in stars
  $\kappa_*/\kappa_\mathrm{tot}$ is indicated by the symbol shape,
  with a legend in the upper left.}
  \label{fig:abs50}
\end{figure}

\begin{figure}
  \centering
  \includegraphics[width=0.45\textwidth]{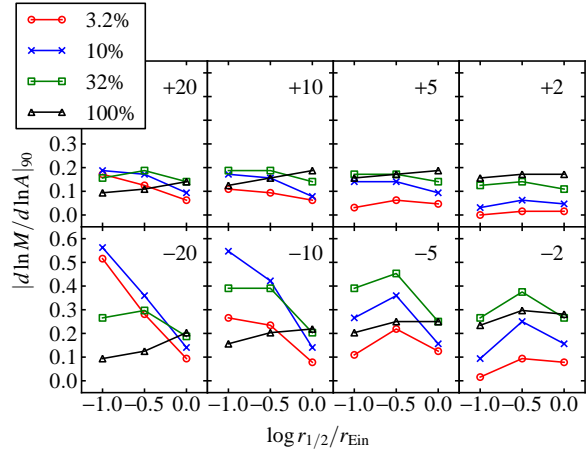}
  \caption{90\% probability bounds on $|\dlmdla|$ as a function of the
  ratio of the source half-light radius to a solar-mass Einstein
  radius. Macro-magnifications are shown in the upper right of each
  plot. The surface mass fraction in stars
  $\kappa_*/\kappa_\mathrm{tot}$ is indicated by the symbol shape,
  with a legend in the upper left. Note the change in vertical scale
  from Figure \ref{fig:abs50}.}
  \label{fig:abs90}
\end{figure}

We created histograms for 8 macro-magnifications, 4 stellar fractions,
and 3 source sizes that are typical of quasars lensed by foreground
galaxies. We simulated finite-sized accretion disks by convolving the
magnification maps with circular Gaussian profiles characterized by a
half-light radius, a parameter which has been shown
\citep{Mortonson:2005p594, Congdon:2007p263} to be more important that
the exact radial profile of the source. Each difference map was
created by subtracting a map which had been convolved with a source
0.1 dex smaller (in each dimension) than the chosen size from a map
convolved with a source 0.1 dex larger. Figure~\ref{fig:histograms}
shows four such histograms, although this is only a fraction of the 96
cases we considered. Figures~\ref{fig:abs50} and \ref{fig:abs90} show
the absolute value of \dlmdla\ bounding 50\% and 90\% of the
probability in each histogram, respectively.

Clearly \dlmdla\ is a small fraction of unity for most magnifications,
stellar fractions, and source sizes. However, there is significant
probability that it will be large enough to have an observable effect,
particularly for small sources in highly-magnified saddle-point
images. We stress, however, that even for large sources, where the
first-order effects of microlensing become very weak, there remains
substantial probability for \dlmdla\ to have significantly nonzero
value.

\subsection{Light curve simulation}
\label{sec:lcsim}

Our derivation in Section~\ref{sec:math} assumed that the time scale
for microlensing was long compared to the time scale for quasar
variability. However, this is not always the case, as quasars can vary
on a range of time scales \citep[e.g.][]{Kelly:2009p895,
Kozlowski:2010p927} and microlens caustic crossings can happen quickly
for smaller source sizes \citep[e.g.][]{Chartas:2009p174}. In order to
test our mathematical description, we ran a more detailed simulation
of the microlensing of a quasar whose area varied linearly with its
flux.

We generated an intrinsic light curve for the source using the damped
random walk stochastic model introduced by \citet{Kelly:2009p895} and
extensively confirmed by \citet{Kozlowski:2010p927} and
\citet{MacLeod:2010p276}. An accretion disk with a Gaussian profile
and a time-averaged half-light radius $10^{-0.2} = 63$\% the size of a
solar-mass Einstein radius was run across two magnification maps, with
a velocity of 1000 km s$^{-1}$ (projected on the source plane). The
maps were chosen to simulate a high-magnification, merging pair of
images with macro-magnifications of $\pm 20$ and stellar mass
fractions $\kappa_*/\kappa_\mathrm{tot} = 0.1$. We simulated 3000
epochs, each separated by a day. At each time step, the area of the
source was adjusted proportionally to the intrinsic light curve, and
the magnification maps were convolved with the resulting source to
find the microlensing magnifications. The simulation produced observed
light curves for the two quasar images which contained the imprint of
the intrinsic quasar light curve as well as microlensing and the
variation of the microlensing magnification with source area
(i.e. $\zeta(t)$).

\begin{figure}
  \centering
  \includegraphics[width=0.45\textwidth]{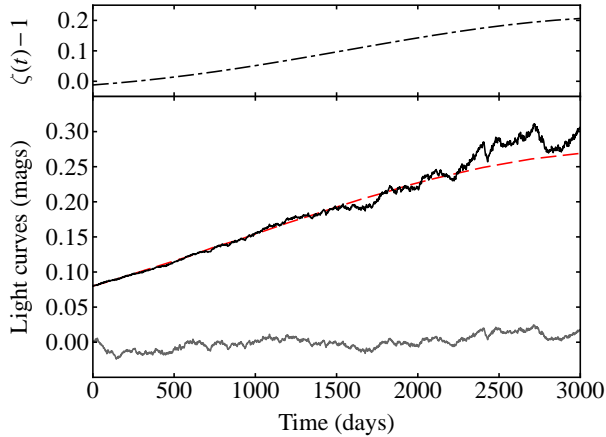}
  \caption{Bottom panel: Simulated difference light curve for a pair
    of highly magnified quasar images. The dashed curve shows the
    difference that would be observed if the disk size was held fixed
    at its average value. The solid black curve includes the effects
    of an accretion disk with a time-varying area.  For comparison, at
    the bottom, we show in gray the input quasar light curve scaled by
    a ``typical'' \dlmdla\ factor of $0.1$ (see
    Figures~\ref{fig:abs50} and \ref{fig:abs90}). Top panel:
    Difference in \dlmdla\ between the two images, as a function of
    time. This is the factor controlling the ``echo'' of the intrinsic
    light curve seen in the difference curve in the bottom panel. The
    source's average half-light radius is 63\% of a solar-mass
    Einstein radius, and the stellar mass fraction is fixed at
    10\%. This simulation assumes perfect correction for the time
    delays between the images.}
  \label{fig:lcsim}
\end{figure}

Figure~\ref{fig:lcsim} shows the difference between the resulting light
curves, including the effect of a time-varying source size, and
compares it to the same quantity with the source size held constant at
the size associated with the mean flux. The first curve shows a clear
echo of the intrinsic source light curve, modulated by $\zeta(t)$ in
good agreement with Equation~\ref{eqn:magdiff}.

\section{Measuring the effect}
\label{sec:epsilon}

We adapted the time-delay fitting algorithm described by
\citet{Kochanek:2006p47} to fit for the presence of $\zeta(t)$ in
measured light curves. This algorithm defines a model light curve
consisting of the sum of a high-order polynomial $s(t)$ (representing
the intrinsic quasar variability) and lower-order polynomials $\Delta
\mu_i(t)$ (representing microlensing fluctuations). The intrinsic
light curve is constrained to be the same for all lensed images,
whereas the microlensing terms are allowed to vary. We modified
Equation 3 of \citet{Kochanek:2006p47} as follows:
\begin{equation}
\label{eqn:newchisq}
\chi^2 = \sum_{i=1}^{N_\mathrm{im}} \sum_{j=1}^{N_\mathrm{obs}}
\left[\frac{m_{ij} - \zeta_i(t_j) s(t_j+\Delta t_i) - \Delta \mu_i(t_j)}{\sigma_{ij}}\right]^2 ~.
\end{equation}
Here $m_{ij}$ and $\sigma_{ij}$ are the data and errors, respectively,
and each $\zeta_i(t)$ is parameterized by a low-order polynomial. (In
practice, $\zeta_i(t)$ is multiplied by only the terms in $s(t)$ that
are higher-order than $N_\mu$, the order of $\Delta \mu_i(t)$, to
avoid degeneracy.) We used a Markov Chain Monte Carlo (MCMC) technique
to minimize $\chi^2$ with respect to the time delays $\Delta t_i$ and
the coefficients parameterizing $\zeta_i(t)$. At each step in the
chain, the linear parameters of $s(t)$ and $\Delta \mu_i(t)$ were
determined by minimizing the $\chi^2$ as in
\citet{Kochanek:2006p47}. We fit the entire time series
simultaneously, as opposed to breaking it into seasons, because we
found that in practice this gave us better control over ``cross-talk''
between the microlensing terms and $\zeta(t)$.

We tested the efficacy of this technique by applying it to simulated
light curves for 30 pairs of images. Like the simulations in
Section~\ref{sec:lcsim}, these curves started with a stochastic
intrinsic light curve, but these did not contain microlensing
variability; we simply multiplied one of each pair of intrinsic light
curves by a slowly-varying curve representing $\zeta(t)$. We deleted
from each light curve 140 out of every 365 points to simulate six
observing seasons, and fit a model to the remaining data to determine
the time delay and $\zeta(t)$. The latter was modeled using Legendre
polynomials to quadratic order: $\zeta(t) = \sum_{i=0}^2 b_i
P_i(\hat{t})$, where $\hat{t}$ is a scaled time variable running from
$-1$ at the beginning of the light curve to $1$ at its end. We used a
quartic ($N_\mu = 4$) to fit the microlensing component, and set the
parameter $\lambda$ to 1.0. We varied the order of the source
polynomial $N_\mathrm{src}$ between 120 and 190, and found a small
decreasing trend in the average $\chi^2$ per degree of freedom for
higher orders. However, the decrease was less than the scatter in the
values, so we concluded that $N_\mathrm{src} = 120$ was sufficient.
On average, the algorithm successfully recovered the $b_i$ and time
delay parameters. Table~\ref{tab:simparams} compares the mean of the
30 best-fit values to the true parameter values, and gives the scatter
in the best-fit values and their typical formal uncertainty.

\begin{deluxetable}{lcccc}
\tablewidth{0pt}
\tablecaption{Best-fit simulated $\zeta(t)$ parameters
  \label{tab:simparams}}
\tablehead{
  &
  \colhead{$b_0$} & 
  \colhead{$b_1$} & 
  \colhead{$b_2$} & 
  \colhead{$\Delta t$ (days)}}
\startdata
True value          & $1.10$ & $0.11$ & $-0.05$     & $\phpm0.00$ \\
Mean                & $1.11$ & $0.11$ & $-0.05$     &     $-0.07$ \\
Standard deviation  & $0.02$ & $0.05$ & $\phpm0.08$ & $\phpm0.66$ \\
Typical uncertainty & $0.02$ & $0.05$ & $\phpm0.06$ & $\phpm0.42$ 
\enddata
\tablecomments{Errors represent formal 68\% ($\Delta \chi^2 = 1$)
confidence intervals.}
\end{deluxetable}

These tests demonstrate that measurements of \dlmdla\ in real-world
lensed quasars have the potential to constrain the value of
$\epsilon$, that is, the relationship between accretion disk size and
luminosity. If the values that are measured (e.g. using the above
method) fall in significantly unlikely regions of the theoretical
histograms (see Section~\ref{sec:hist}), the $\epsilon = 1$
hypothesis, that is, a blackbody with power-law temperature profile,
will be called into question.

The presence of contaminating light from non-disk sources in our
quasar image light curves acts as a spurious offset in the measured
$\zeta(t)$ curves (see Equation~\ref{eqn:magdiff}). It must be
controlled as much as possible in order to make robust estimates of
the value of $\epsilon$.  Contamination from the lens galaxy or from
the quasar's host galaxy lensed into an Einstein ring can be measured
and removed using high-quality ground-based or space-based imaging
contemporaneous with some part of the light curve. Spectroscopic
methods are necessary to remove contamination from non-disk regions of
the quasar, such as emission lines or the dusty torus. However, the
contamination from these sources will largely cancel in the absence of
strong microlensing of these regions.

\section{Application to \hefour}
\label{sec:hefour}

\begin{figure*}
  \centering
  \includegraphics[width=0.68\textwidth]{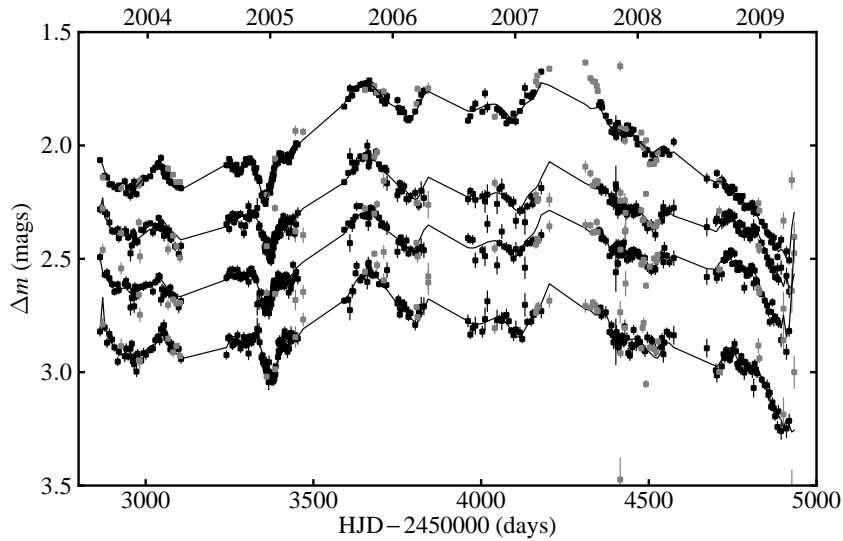}
  \caption{Light curves of images A, B, C, and D (downward from top)
  of \hefour. For clarity, offsets of $-0.35$ and $-0.1$ magnitudes
  have been applied to images B and C, respectively. The solid lines
  are the best-fitting model light curves. The gray points, indicating
  epochs containing points that differed by more than
  $\sqrt{10}\sigma$ from an initial fit to all the data, were not used
  for the fit.}
  \label{fig:hefourlc}
\end{figure*}

We have applied the adapted fitting algorithm described in
Section~\ref{sec:epsilon} to a six-year $R$-band light curve of the
quadruply lensed quasar \hefour. The 332 epochs of data are the
results of a photometric monitoring campaign between 2003 December and
2009 April. Data at four epochs were obtained with SPICAM on the
3.5-meter telescope of the Apache Point Observatory (APO) in New
Mexico. The rest of the data were obtained with the ANDICAM camera
\citep{DePoy:2003p827} on the 1.3-meter SMARTS telescope. The first
141 epochs of this data set, including all the APO data, were
presented in detail by \citet{Kochanek:2006p47}, and the remainder of
the data were reduced using the procedures described there. We defer
publication of the remainder of the data, as well as a detailed study
of the time delays and lens models, to a later study, as they are not
our present focus. However, we show the light curves in
Figure~\ref{fig:hefourlc}. The gray points indicate data from epochs
that were excluded from our fit because one or more measurements
differed from an initial fit to all the data enough to contribute more
than $10$ to the total $\chi^2$. A microlensing event is visible in
the light curve of image A between seasons 4 and 5; because our
microlensing model is a low-order polynomial incapable of reproducing
such fine structure, the test excludes some points in these
seasons. In hindsight, it makes sense to exclude these points, since
our current purpose is to detect quasar variability, not microlensing.

We modeled $\zeta_i(t)$ for images A, C, and D as linear functions of
time (image B was used as the reference image). For each
(non-reference) image, the curve was parameterized by $\zeta_i(t) =
b_{0i} + b_{1i} \hat{t}$, where $\hat{t}$ is again defined as a scaled
time variable running from $-1$ at the beginning of the light curve to
$1$ at its end. This choice allowed us to test for the presence of a
\dlmdla\ term even in the presence of contamination of the quasar
light curves because the contamination $C$ should not vary with
time. The source light curve and microlensing curves were modeled as
polynomials of order $N_\mathrm{src} = 120$ and $N_\mu = 4$,
respectively. Including $\zeta(t)$ dramatically improves the fit to
the data from $\chi^2=2036$ for $N_\mathrm{dof}=1190$ to $\chi^2=1699$
for $N_\mathrm{dof}=1184$. We show the best-fit model light curves
along with the data in Figure~\ref{fig:hefourlc}, and the best-fit
parameters for the $\zeta_i(t)$ curves in
Table~\ref{tab:zetaparams}. The best-fit time delays are the same
within the errors whether or not $\zeta(t)$ is included, and are
broadly consistent with those reported by \citet{Kochanek:2006p47}, in
the sense that the order of arrival is preserved. In detail, our
best-fit time delays differ from those previously reported by $\sim 1$
to $4\sigma$ (though correlations between the measured values reduce
this tension). The discrepancy is due to our strategy of modeling the
microlensing as a single polynomial spanning all six seasons, instead
of a separate polynomial for each season, as was done in the earlier
work. When we fit each season separately, we recovered time delays
within $1\sigma$ of the previously reported values. Increasing $N_\mu$
improves the agreement between our delays and the reported delays, but
the improvement is rather slow; our tests required $N_\mu \sim 20$
before the discrepancy was cut in half. It is not our present goal to
recalculate the time delays, but we discuss in
Section~\ref{sec:conclusions} some potential implications of our
analysis for time delay estimates. We performed a second fit, holding
the time delays constant at the previously reported values, and found
that the $\chi^2/N_\mathrm{dof}$ increased to $1750/1187$, while the
best-fit $\zeta(t)$ parameters were virtually unchanged. We report
them in Table~\ref{tab:zetaparams}.

The significantly non-zero linear coefficient $b_1$ in the best-fit
$\zeta_i(t)$ curves for image A (see Table~\ref{tab:zetaparams})
suggests that the changing accretion disk size has an appreciable
effect on at least that image's light curve. It is worth noting that
this image was already thought to be affected by microlensing
\citep{Kochanek:2006p47}. However, the limitations associated with our
analysis method prevent us from placing too much quantitative
confidence in this result. The fit requires that the microlensing
variability, as well as the variation in $\zeta(t)$, be slow in
time. It also did not give sensible results when we attempted to fit
individual seasons because of instabilities between the microlensing
terms $\Delta \mu$ and $\zeta(t)$. Finally, it takes no account of the
correlation between the microlensing caustic pattern and its
derivative with respect to source size --- the microlensing terms and
the $\zeta(t)$ terms were fit independently. Because of these
limitations, and because we have analyzed only a single lensed quasar
without carefully addressing the problem of contaminating light, we
cannot yet comment on the value of $\epsilon$ in detail.

The Bayesian Monte Carlo analysis method described by
\citet{Kochanek:2004p58} and updated to include stellar motions by
\citet{Poindexter:2009p3213} addresses most of these limitations. This
method simulates microlensing light curves using the same technique as
described in Section~\ref{sec:simulations}, but on a much larger
scale. Simulated light curves that are found to be consistent with the
measured data contribute to a Bayesian determination of parameters
such as the source size and average microlens mass. The $\epsilon$
parameter may be added to the analysis as an additional parameter, and
the source size allowed to vary with time when the light curves are
simulated. This treatment allows for arbitrary microlensing
variability, and would treat the \dlmdla\ terms self-consistently. We
are optimistic that this improved analysis, applied to a number of
light curves, will allow constraints to be placed on $\epsilon$.

\begin{deluxetable}{ccccc}
\tablewidth{0pt}
\tablecaption{\hefour\ best-fit $\zeta(t)$ parameters
  \label{tab:zetaparams}}
\tablehead{
  &
  \multicolumn{2}{c}{Fixed delays} &
  \multicolumn{2}{c}{Free delays} \\
  & 
  \colhead{$b_0$} & 
  \colhead{$b_1$} & 
  \colhead{$b_0$} & 
  \colhead{$b_1$}}
\startdata
A & $0.92\err{.04}{.03}$ &     $-0.45\err{.07}{.06}$ & $0.93\err{.03}{.04}$ &     $-0.43\err{.06}{.07}$ \\
B &           $\equiv 1$ &                $\equiv 0$ &           $\equiv 1$ &                $\equiv 0$ \\
C & $0.98\err{.05}{.03}$ & $\phpm0.11\err{.10}{.07}$ & $1.00\err{.04}{.04}$ & $\phpm0.12\err{.09}{.08}$ \\
D & $1.15\err{.06}{.04}$ &     $-0.10\err{.11}{.09}$ & $1.16\err{.05}{.05}$ &     $-0.08\err{.09}{.10}$ \\
$\chi^2/N_\mathrm{dof}$ & \multicolumn{2}{c}{$1750/1187$} & \multicolumn{2}{c}{$1699/1184$} 
\enddata
\tablecomments{Errors represent formal 95\% ($\Delta \chi^2 = 4$)
confidence intervals. Because image B was the reference image, its
parameters were fixed at the nominal values.}
\end{deluxetable}

\section{Conclusions}
\label{sec:conclusions}

We have described a new effect arising from the microlensing of quasar
accretion disks whose sizes vary in time. Assuming the change in size
is proportional to some power of the change in luminosity of the
quasar, we have derived what the effect would be on the light curves
of lensed quasar images (see, e.g., Equation~\ref{eqn:magdiff}). The
effect arises from the dependence of the microlensing magnification on
the area of the source. This logarithmic derivative, \dlmdla, depends
on the macro-magnification of the image, the surface density of
microlens stars, and the size of the source, along with the detailed
arrangement of the microlensing caustic pattern. We have constructed
probability distributions for the value of \dlmdla\ for various
parameter combinations, and find that its value can be significantly
nonzero. In particular, we note that it does not decrease in magnitude
with increasing source size as strongly as the microlensing
magnification itself does. We have also checked our mathematical model
by constructing simulated light curves which demonstrate the
microlensing of a time-varying disk.

We have demonstrated how fits may be made to such light curves to
recover $\zeta(t)$, defined as the difference in \dlmdla\ between
pairs of images. Some systematic errors are introduced by the presence
of contaminating flux in the light curves, and we have discussed how
these may be controlled. Applying the fitting method to the light
curves of the four images of the lensed quasar \hefour, we find
evidence for significant variation in this difference for image A
relative to image B, indicating that the effect is likely occurring
for this image. However, limitations of this fitting method, in
particular the inconsistent treatment of the correlation between the
microlensing and its size derivative, prevent us from saying anything
quantitative. We propose, but have not implemented, an improved
treatment based on the Bayesian Monte Carlo method of
\citet{Kochanek:2004p58}. This method is fully self-consistent, and
ought to be more robust. The one limitation of this approach is that
an a priori model of the source variability will have to be used to
control the source size rather than the best fit estimate of the
source light curve generated for each trial. A determination of
$\zeta(t)$, in particular for a large sample of lensed quasar images,
may be used to test the power-law slope of the dependence of quasar
luminosity on disk size, a quantity that is predicted to be unity for
simple blackbody disk models.

Our estimates of the time delays of \hefour\ are only roughly
consistent with those of \citet{Kochanek:2006p47}, whether or not we
include the effect of $\zeta(t)$. Attempts to use the time delays of
lensed quasar images for precision cosmology have often been hampered
by a lack of robustness in measured delays, though the more serious
problem remains model degeneracies \citep[see discussion
in][]{Kochanek:2006p91}. In our case the tension is at least partially
due to our choice of a low-order polynomial stretching across the
entire 6-year light curve to model the microlensing
variability. Although it was necessary for the sake of numerical
stability, this model lacks the freedom necessary to describe the
highest-frequency microlensing structures. This is another reason to
adopt the self-consistent method described above. It is possible that
previously unmodeled variable source size effects are also behind some
of the discrepancy, in which case including the $\zeta(t)$ term would
result in more accurate delay estimates. However, we note that we did
not find significant differences in the best-fit time delays of
\hefour\ when the $\zeta(t)$ terms were allowed to vary versus when
they were fixed at nominal values, nor did the best-fit $\zeta(t)$
terms change when the time delays were fixed at the previously
reported values, so this remains speculation at best.

\acknowledgements
This research was supported by NSF grant AST-0708082.

\bibliography{ms}

\end{document}